\documentclass[fleqn,10pt]{wlscirep}
\usepackage[utf8]{inputenc}
\usepackage[T1]{fontenc}
\usepackage{multirow}
\title{A curated dataset for data-driven turbulence modelling}

\author[1*]{Ryley McConkey}
\author[1$\dag$]{Eugene Yee}
\author[1$\dag$]{Fue-Sang Lien}
\affil[1]{University of Waterloo, Department of Mechanical and Mechatronics Engineering, 200 University Avenue, Waterloo, ON, N2L 3G1, Canada}

\affil[*]{Corresponding author. \textit{Email address:} rmcconke@uwaterloo.ca.}

\affil[$\dag$]{These authors jointly supervised this work.}

\begin{abstract}

The recent surge in machine learning augmented turbulence modelling is a promising approach for addressing the limitations of Reynolds-averaged Navier-Stokes (RANS) models. This work presents the development of the first open-source dataset, curated and structured for immediate use in machine learning augmented turbulence closure modelling. The dataset features a variety of RANS simulations with matching direct numerical simulation (DNS) and large-eddy simulation (LES) data. Four turbulence models are selected to form the initial dataset: $k$-$\varepsilon$, $k$-$\varepsilon$-$\phi_t$-$f$, $k$-$\omega$, and $k$-$\omega$ SST. The dataset consists of 29 cases per turbulence model, for several parametrically sweeping reference DNS/LES cases: periodic hills, square duct, parametric bumps, converging-diverging channel, and a curved backward-facing step. At each of the 895,640 points, various RANS features with DNS/LES labels are available. The feature set includes quantities used in current state-of-the-art models, and additional fields which enable the generation of new feature sets. The dataset reduces effort required to train, test, and benchmark new models. The dataset is available at \url{https://doi.org/10.34740/kaggle/dsv/2044393}.

\end{abstract}
\begin{document}

\flushbottom
\maketitle

\thispagestyle{empty}


\section*{Background \& Summary}


Numerical simulations in weather forecasting, wind and hydroelectric energy, aerospace vehicle design, automotive design, turbomachinery, nuclear plant design, and many other applications all rely on closure models to accelerate simulations while modelling the complex physical phenomenon of turbulence. While higher resolution techniques such as large-eddy simulation (LES) and direct numerical simulation (DNS) are becoming more widespread, the computational demands compared to current capabilities make these techniques unaffordable for many industrial simulations. For this reason, Reynolds-averaged Navier-Stokes (RANS) simulations are expected to remain the dominant tool for predicting flows of practical relevance to engineering and industrial problems over the next few decades \cite{CFD2030}. However, flows with strong adverse pressure gradients \cite{Wilcox1994}, separation \cite{Catalano2003}, streamline curvature \cite{Pope1975}, and reacting chemistry are often poorly predicted by RANS approaches. Developing methods to improve the accuracy of RANS simulations will help bridge this critical capability gap between RANS and LES\cite{Witherden2017}. 

Several recent investigations have demonstrated the potential of applying machine learning to the development of turbulence closure models for RANS. Ling et al. \cite{Ling2016} constructed a tensor basis neural network (TBNN), which predicts the anisotropy tensor using five invariant scalars derived from the mean strain and rotation rate tensors. The TBNN turbulence closure model developed by Ling et al. \cite{Ling2016} is effectively a fifth-order eddy viscosity model, with locally varying coefficients predicted via deep learning. The ability to express such a locally-tuned, high-order relationship between the strain rate and anisotropy tensors is a powerful method to improve the accuracy of RANS simulations. Wu et al. \cite{Wu2018} developed a random-forests-based model, which directly predicts the Reynolds stress anisotropy. Kaandorp \cite{Kaandorp2018} and Kaandorp and Dwight \cite{Kaandorp2020}  proposed a tensor basis random forest (TBRF) model, which is the random forests analogue to the TBNN proposed by Ling et al.~\cite{Ling2016}. While the different models by Ling et al.~\cite{Ling2016}, Wu et al.~\cite{Wu2018}, Kaandorp and Dwight~\cite{Kaandorp2020}, Zhu and Dinh~\cite{Zhu2020}, Zhang et al.~\cite{Zhang2019a}, Fang et al.~\cite{Fang2019}, and Song et al.~\cite{Song2019} all show promise, the results cannot be directly compared --- each investigation used a different set of input features and labels, with different numerical settings chosen for feature generation. For this reason, Duraisamy \cite{Duraisamy2020} recently highlighted the need for a benchmark dataset for machine-learnt closure models.

To generate a set of input features, the current requirement is for every investigator to generate a set of RANS simulations that match the DNS/LES reference cases. This requirement has several drawbacks. As the number of included datasets grows, the effort required grows. The development of the ImageNet dataset spurred rapid growth of the computer vision field, which would not have been possible otherwise. From an effort point of view, the availability of a curated dataset dramatically increases the time spent developing the models themselves, rather than setting up many RANS simulations to gather input features. 

Another major drawback of the current approach arises from the issue of reproducibility in the field of computational fluid dynamics (CFD). Often, CFD studies are difficult to reproduce, due to a large number of input conditions \cite{Mesnard2017}. Each investigation will use different meshes, numerical schemes, turbulence models, and other selections which affect the solution. The field of machine learning has also been plagued with reproducibility challenges, even with the widespread use of benchmark datasets \cite{Pineau2020}. While machine-learnt turbulence models are a promising approach, the development of these models could be significantly impeded by mixing two fields where reproducibility is a challenge. A well-documented, widely available dataset solves at least one aspect of the reproducibility issue, in that all models can at least be trained in the same environment, using the same input features and labels.

Motivated by the lack of a sufficient dataset, the present work aims to develop a set of RANS simulations of highly resolved reference cases in order to generate a curated dataset\cite{McConkey2021}. In this work, the numerical methods for the RANS simulations are presented, along with the selection and calculation of the input features for machine learning models. In doing so, the present work aims to present a large computational dataset, curated and logically structured for immediate use in developing next-generation turbulence closure models for RANS using data-driven machine learning. Table \ref{tbl:inputsoutputs} summarizes the inputs and outputs of the present work.

\section*{Methods}





 

\subsection*{Description of reference cases} \label{sec:description}
An important aspect of dataset selection for data-driven turbulence modelling is sweeping of a parameter space. A deep insight followed by a deeper understanding of the fluid phenomena can be obtained by providing information on how the geometry and/or the Reynolds number changes the flow behaviour. In contrast, single-point measurements are only valuable in approximating a universal mapping between inputs and outputs. The majority of the datasets used here involve sweeping through some parameter space. Table \ref{tbl:cases} summarizes the cases used in the dataset.

\subsubsection*{Periodic hills}
\begin{figure}[h]
    \centering
    \includegraphics[]{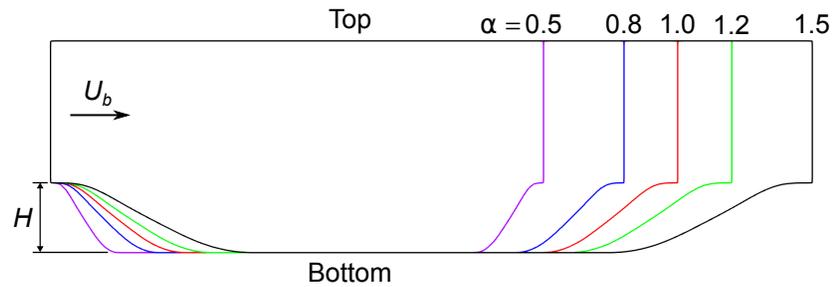}
    \caption{The geometry for the five periodic hills cases. Further detail is given in Xiao et al.~\cite{Xiao2020}. The Reynolds number for this case is calculated based on the hill height $H$ and mean bulk velocity $U_b$. These parameters are fixed for all cases, so $Re_H$ remains fixed at 5,600.}
    \label{fig:periodichills}
\end{figure}

Flow over periodic hills with cyclic boundary conditions is a common benchmark problem for turbulence modelling. The periodic hills case features separation, an important phenomenon for RANS models to accurately capture due to the prominence of strongly separated flows in many industrial settings. To provide a parameterized dataset for data-driven turbulence modelling, Xiao et al.~\cite{Xiao2020} performed DNS of flow over a series of periodic hills. This dataset consists of five cases, characterized by the steepness ratio $\alpha$. The values of $\alpha$ selected are $\alpha=0.5, 0.8, 1.0, 1.2, $ and 1.5, which results in a range of separated flows. The geometry for the five periodic hills cases is shown in Figure \ref{fig:periodichills}. The Reynolds number based on bulk velocity and crest height for all cases is fixed at $Re=5,600$.

\subsubsection*{Square duct}
\begin{figure}[h]
    \centering
    \includegraphics[]{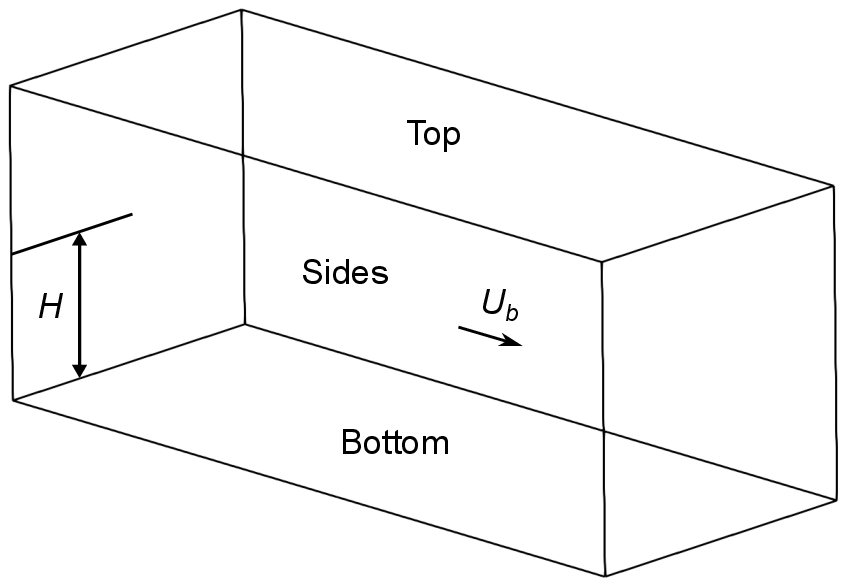}
    \caption{The geometry for the square duct cases. The cases vary by changing the Reynolds number from 1,100 to 3,500, which is calculated based on the duct half-width $H$ and mean bulk velocity $U_b$.}
    \label{fig:duct}
\end{figure}

The DNS dataset for flow in a square duct by Pinelli et al.~\cite{Pinelli2010} has been widely used in data-driven turbulence modelling. This dataset consists of 16 cases, all with the same fixed geometry shown in Figure \ref{fig:duct}. The Reynolds number based on the duct half-width varies between 1,100 and 3,500. The flow in a square duct is a challenging test case for eddy viscosity models. Linear eddy viscosity models are unable to predict the secondary corner vortices which form in the duct. These structures are Prandtl's secondary motion of the second kind \cite{Pinelli2010}. The dataset contains the mean velocities and Reynolds stresses in a cross-section of the duct. The inclusion of this dataset allows the machine-learnt model to incorporate the Reynolds number dependence of these challenging secondary motions, from the transitional to the fully turbulent regimes. Additionally, it is the only three-dimensional (3D) flow in the dataset, for which the Reynolds shear stresses $\overline{u'w'}$ and $\overline{v'w'}$ are nonzero.

\subsubsection*{Parametric bumps}
\begin{figure}
    \centering
    \includegraphics[]{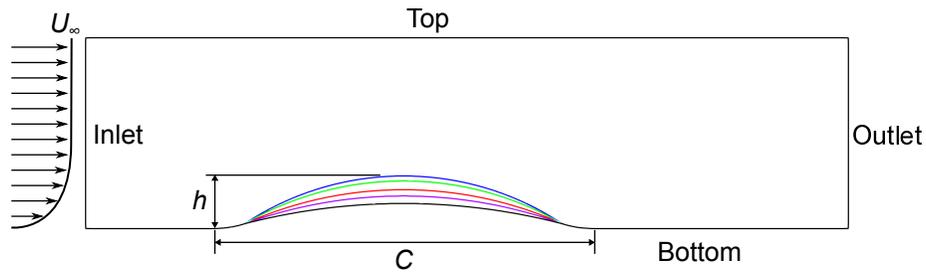}
    \caption{The geometry for the five parametric bump cases. The bump length $C$ is fixed at 305 mm, and the bump height varies as $h$ = 20, 26, 31, 38, and 42 mm. Further detail is given in Matai and Durbin \cite{Matai2019a}. The Reynolds number based on maximum inlet velocity and step height varies from $Re_h=13,260$ to $Re_h=27,850$, with the momentum thickness Reynolds number fixed at $Re_\theta=2,500$.}
    \label{fig:bump}
\end{figure}

The LES dataset for flow over a family of bumps by Matai and Durbin~\cite{Matai2019a} has been recently made available for data-driven closures. The bump is a circular arc, with convex fillets on either end. The dataset is characterized by the bump height $h$, which is the highest point of the circular arc as shown in Figure \ref{fig:bump}. The Reynolds number based on momentum thickness and inlet free stream velocity $U_\infty$ is fixed at $Re_\theta=2,500$, while the Reynolds number based on bump height and $U_\infty$ varies from $Re_h\approx13,250$ to $27,850$. At $h=20$ mm, the flow remains attached along the bump, while increasing the height further results in slight separation at $h=26$ mm. For the highest bump corresponding to $h=42$ mm, a small separated region forms behind the bump. While the periodic hills dataset features massively separated flows, the bump cases incorporate a smaller degree of separation. Matai and Durbin found that the mild separation causes a high turbulent kinetic energy (TKE) zone to depart from the bump ahead of the separated region, which is not the case for massively separated flows. Matai and Durbin attributed this region to the adverse pressure gradient generating a mean shear profile. Another important effect captured in the parametric bump case is strong disequilibrium. The parametric bump dataset is highly valuable for training machine-learnt closure models due to the high Reynolds number, parametrically sweeping geometry, physics unique to mildly separated flows, and strong disequilibrium.

\subsubsection*{Converging-diverging channel}
\begin{figure}
    \centering
    \includegraphics[]{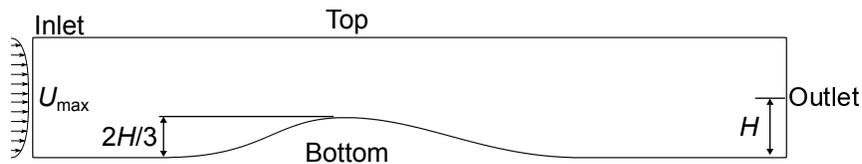}
    \caption{The geometry for the two converging-diverging channel cases, corresponding to Reynolds numbers of $Re_H=12,600$ and 20,580. The Reynolds number for these two cases is based on the channel half height $H$ and the maximum inlet velocity $U_\text{max}$.}
    \label{fig:cndv}
\end{figure}

Two datasets are available for flow over an identical converging-diverging geometry at $Re_H=12,600$ and $Re_H=20,580$, shown in Figure \ref{fig:cndv}. The Reynolds number for this case is based on the maximum inlet velocity and the channel half-height $H$. The lower Reynolds number dataset comes from the DNS by Laval and Marquielle~\cite{Laval2011}, and Marquillie et al.~\cite{Marquillie2008}. The higher Reynolds number dataset was generated by Schiavo et al.~\cite{Schiavo2015} using LES. The bump height is approximately $2H/3$. A fully developed internal channel flow enters the domain and impinges on the abrupt upstream side of the bump. The flow accelerates as the channel converges, then decelerates over the gradual downstream side of the bump. At $Re_H=12,600$, a thin separation bubble forms along the downstream slope. Along the flat upper wall, the flow remains attached but on the cusp of separation. At $Re_H=20,580$, the separation bubble grows. The cases contain valuable information about the Reynolds number effect on separation, reattachment, and development of a turbulent boundary layer under an adverse pressure gradient. The long domain downstream of the bump for $Re_H=20,580$ effectively provides an additional set of LES information for developing plane channel flow.

\subsubsection*{Curved backward-facing step}
\begin{figure}
    \centering
    \includegraphics[]{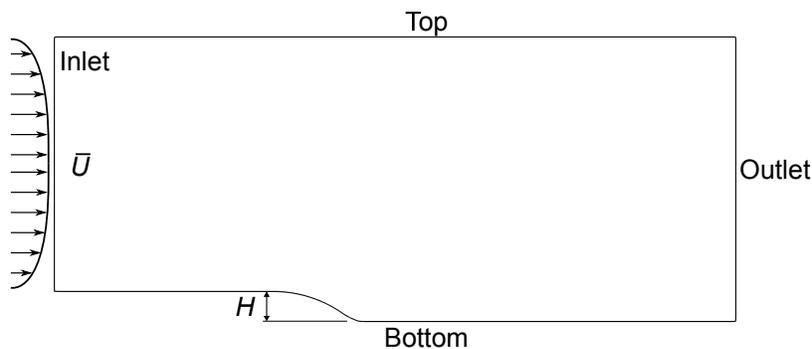}
    \caption{The geometry for the curved backward-facing step case. The Reynolds number $Re_H=13,700$ is based on the mean inlet velocity $\overline{U}$ and the step height $H$.}
    \label{fig:cbfs}
\end{figure}

The curved backward-facing step case simulated by Bentaleb et al.~\cite{Bentaleb2012} using LES was also included in the dataset. The geometry for this case is shown in Figure \ref{fig:cbfs}. While this is the only case that does not feature parametric variation, it contains an additional set of data on separation and reattachment. While other cases in the dataset feature separation after an acceleration of the flow, the curved backward-facing step case features separation of a fully developed turbulent boundary layer. This phenomenon is difficult for RANS models to predict, and therefore the LES results were included in the dataset. While the original work by Bentaleb et al.~\cite{Bentaleb2012} defined a Reynolds number based on the maximum inlet velocity, we found that the large channel height meant that the mean velocity for all turbulence models was within 10\% of the maximum velocity, so to approximate the reference case, defining the Reynolds number for the dataset based on the mean inlet velocity was sufficient.

\subsection*{Computational method}\label{sec:computational}
The flow is assumed to be incompressible, viscous, steady, and turbulent for all cases. Under these conditions, the fluid properties are specified by the kinematic molecular viscosity $\nu$. Table \ref{tbl:viscosity} summarizes the viscosity used for each case.

The open-source library OpenFOAM v2006 \cite{Openfoam} was used to generate the dataset. The ability to replicate CFD is greatly improved by supplying the mesh and settings files \cite{Mesnard2017}. The dataset includes the OpenFOAM case files, including the meshes used for all the cases, and the full details of the settings used. Supplying the OpenFOAM files also reduces the effort required for \textit{a posteriori} testing. This practice is following Xiao et al.~\cite{Xiao2020}, who included the OpenFOAM files with their dataset. While this section highlights the basic numerical settings used, the reader is referred to the dataset for the complete OpenFOAM settings.

\subsubsection*{Numerical schemes}
A standardized set of numerical schemes was used for all cases. The numerical schemes represent commonly used RANS schemes, which represent a good trade-off between stability and accuracy. For discretizing the convective terms in the momentum equations, a second-order upwind scheme was used. For discretizing convective terms in the turbulence transport equations, a first-order upwind scheme was used. For the diffusive terms, a second-order central difference scheme was used. Since all the flow cases are steady, the transient terms were set to zero.

The simpleFoam solver was used to solve the equations iteratively. The semi-implicit method for pressure-linked equations-consistent (SIMPLEC) algorithm was used to accelerate convergence. For some cases, additional non-orthogonality correcting loops were applied to the pressure equation. The generalized geometric algebraic multigrid (GAMG) solver was used for the pressure equation, and the preconditioned bi-conjugate gradient (PBiCGStab) solver was used for all other equations. In all cases, the residuals for all flow variables converged below $10^{-6}$. Residual plots for all simulations are provided in the dataset.

\subsubsection*{Turbulence modelling}
The two most common families of turbulence closure models, $k$-$\varepsilon$ and $k$-$\omega$, include many sub-models. Previous investigations on machine-learnt models for predicting the anisotropy tensor have augmented the standard $k$-$\varepsilon$ model\cite{Ling2016}, the Launder-Sharma low Reynolds number $k$-$\varepsilon$ model \cite{Kaandorp2018}, and the $k$-$\omega$ model \cite{Wu2018, Kaandorp2020}. Four representative turbulence models were selected for the dataset: namely, the standard $k$-$\varepsilon$ \cite{Launder1974}, $k$-$\varepsilon$-$\phi_t$-$f$ \cite{Laurence2005}, $k$-$\omega$, and the $k$-$\omega$ shear stress transport (SST) \cite{Menter2003} turbulence closure models. In this work, $\phi_t$ is used to denote the anisotropy measure $\overline{v'^2}/k$ to align with the variable naming in OpenFOAM. Here, $\overline{v'^2}$ denotes the wall-normal Reynolds stress. The default coefficients were used for all turbulence models \cite{Openfoam}.

The $k$-$\varepsilon$-$\phi_t$-$f$ model is a more sophisticated model than the $k$-$\varepsilon$ and $k$-$\omega$ models, through the inclusion of an additional transport equation for the anisotropy measure $\phi_t\equiv\overline{v'^2}/k$, and an elliptic equation for $f$. $f$ is a scalar which predicts TKE redistribution from the streamwise to the wall-normal Reynolds stress. This model is an improved version of the original $\overline{v'^2}$-$f$ model proposed by Durbin \cite{Durbin1991}, and the improved "code-friendly" version developed by Lien and Kalitzin~\cite{Lien2001}. The additional quantities enable the creation of new input features not available in the previous two-equation investigations. Both additional scalars satisfy all desired invariance properties, including Galilean invariance.

For all turbulence models, the mesh was sufficient for a low Reynolds number wall treatment. Low Reynolds number wall boundary conditions are provided for $k,\varepsilon,\omega,$ and the eddy viscosity $\nu_t$ in OpenFOAM \cite{Liu2016}. A fixed-value $k=0$ boundary condition was applied at no-slip walls. At no-slip walls, the following low Reynolds number fixed value boundary condition was applied for $\varepsilon$:  
\begin{equation}
    \varepsilon=\varepsilon_{vis} = 2wk \frac{\nu}{y^2}\ ,
\end{equation}
where $w$ are the cell corner weights\cite{Openfoam}.
For $\omega$ the following fixed value boundary condition was applied at no-slip walls:

\begin{equation}
    \omega = \frac{6 \nu}{\beta_1 y^2}\ ,
\end{equation}
where $\beta_1 =0.075$.

\subsubsection*{Domain and boundary conditions}
The domain and boundary conditions for all cases were selected to match the DNS or LES reference simulations. There are two main types of boundary conditions used in the dataset: fixed-free, and streamwise cyclic. While the periodic hills and duct cases utilize a streamwise cyclic boundary condition, the bump, converging-diverging channel, and curved-backward facing step cases employ a fully-developed inlet velocity profile, and a zero-gradient outlet. The simulations here involve four different turbulence models, each with different fields. The units used for each variable are given in Table \ref{tbl:units}.

The geometry for the square duct is shown in Figure \ref{fig:duct}. The dimensions for this 3D case are given in terms of the duct half-width $H$. The duct is a $2H\times 2H\times 5H$ box. Wall boundary conditions were applied for the top, bottom and sides of the duct. The boundary conditions for the square duct case are summarized in Table \ref{tbl:duct}.

The periodic hills case is a two-dimensional (2D) flow, with the domain geometry characterized in terms of the hill height $H$, as shown in Figure \ref{fig:periodichills}. The domain height is fixed at $3.04 H$, and the domain width changes from $7.07H$ to $10.9H$, as the parameter $\alpha$ changes. The boundary conditions for the periodic hills case are identical to the duct case (see Table \ref{tbl:duct}), with streamwise cyclic boundary conditions applied for all flow variables. Both the top and bottom boundaries are treated as no-slip walls. To maintain a constant bulk velocity in the flow, a mean pressure gradient source term is added to the momentum equation. Therefore, the pressure field for cases with cyclic boundary conditions should be interpreted as the deviation from the mean pressure field.

For the parametric bump, converging-diverging channel, and curved backward-facing step cases, the DNS and LES simulations utilized a fully-developed inlet flow generated by a "feeder" simulation. To generate the RANS inlet condition, a similar approach to the DNS and LES was taken: a flat version of the domain was simulated with fixed-free boundary conditions to allow the flow to fully develop before entering the domain of interest. Equations for isotropic turbulence are commonly used to estimate the RANS boundary conditions for fixed turbulence inlets. For the feeder simulations, the following equations were used to estimate turbulence quantities at the inlet:

\begin{equation}\label{eq:k}
k = \frac{3}{2}(UI)^2\ ,
\end{equation}

\begin{equation}\label{eq:epsilon}
    \varepsilon = C_\mu^{3/4} \frac{k^{3/2}}{L_t}\ ,
\end{equation}

\begin{equation}\label{eq:omega}
    \omega = \frac{\varepsilon}{0.09 k}\ ,
\end{equation}
where $I$ is the turbulent intensity, $L_t$ is the turbulence length scale and $C_\mu$ is a turbulence closure coefficient.

The parametric bumps case is unique in this dataset in that the top boundary is zero-gradient, compared to the walls used in the other cases. The inlet free-stream velocity $U_\infty$ for the LES reference simulation was 16.77 m/s. To recreate these conditions, the inlet boundary conditions for the flat cases were adjusted to produce $U_\infty=16.77$ m/s. It should be noted that this is an approximation of the LES inlet condition, because the four different turbulence models all produce different $U_\infty$. For the dataset, the mean velocity used for all turbulence models was the same (Table \ref{tbl:bump_develop}), so that the boundary conditions are comparable between turbulence models. The boundary conditions for generating a fully-developed inlet profile for the bump case are summarized in Table \ref{tbl:bump_develop}. After generating a fully-developed profile, the $U,k,\varepsilon,$ and $\omega$ fields were used as fixed-value inlet conditions for the bump cases. The boundary conditions for the bump cases are summarized in Table \ref{tbl:bump}. The domain size for the parametric bump set is fixed at $1.33 H \times 0.5 H$.

A similar procedure for the bump case was completed to generate inlet conditions for the converging-diverging channel and curved-backward facing step cases. However, for the latter two cases, the top boundary is a wall. The boundary conditions for the converging-diverging channel case were adjusted to produce a maximum velocity of $U_\text{max}=1.0$ m/s, to match the reference simulations. Similarly, for the curved backward-facing step case, the mean velocity was set to $1.0$ m/s to match the reference simulation. The boundary conditions for the flat, developing flow cases are shown in Tables \ref{tbl:cndv_develop} and \ref{tbl:cbfs_develop}, and the boundary conditions for the cases in the data set are shown in Tables \ref{tbl:cndv} and \ref{tbl:cbfs}. The domain size for the $Re_H=12,600$ converging-diverging channel is $12.6H \times 2H$, while for $Re_H=20,580$ the domain is enlarged to $25.3H \times 2H$ by extending the outlet length. The curved backward-facing step domain is $22.7H\times 9.48 H$.

\subsubsection*{Mesh} 
OpenFOAM's utilities were used to generate the meshes. The mesh generation method varied from case to case, as some cases have changing geometries. Table \ref{tbl:mesh} summarizes the meshes used. All meshes met the low Reynolds number wall treatment criterion of $y^+\approx1$ or below. Here, $y^+ \equiv u_\tau y_w/\nu$ is the normalized wall-normal distance, where $y_w$ is the wall-normal distance, and $u_\tau$ is the wall friction velocity.
In all cases, the mesh was either hexahedral or hexahedral-dominant. A high-quality mesh is important for generating input features for machine learning, in that some terms involve terms that are sensitive to the mesh quality. For example, the basis tensor $\hat{\mathcal{T}}_{10}$ in a general representation of the Reynolds stress tensor proposed by Pope \cite{Pope1975} is fifth order in terms of the velocity gradient tensor. In developing the feature set here, we found that to keep these terms stable, the number of tetrahedral cells in the domain must be minimized. However, many industrial meshes contain tetrahedral cells, and are of poorer quality than the structured meshes generated here. While CFD results are normally sensitive to the mesh used, machine learning models are especially sensitive to the mesh quality. Poorer meshes result in increased noise and more outliers in the input feature set. 

\begin{figure}[]
    \centering
    \includegraphics[width=0.7\textwidth]{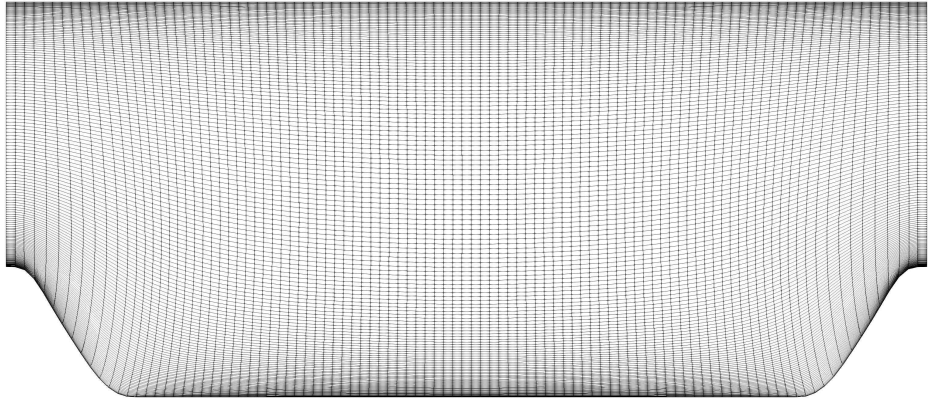}
    \caption{Structured hexahedral mesh used to discretize the $\alpha = 0.5$ periodic hills case.}
    \label{fig:phll_mesh}
\end{figure}

The mesh for the steepest periodic hills case ($\alpha = 0.5$) is shown in Figure \ref{fig:phll_mesh}. The RANS meshes for all periodic hills cases were provided by Xiao et al.~\cite{Xiao2020}. The periodic hills mesh is a structured mesh, with cells concentrated near the boundary layer. While the geometry changes by varying the hill steepness and domain length, the number of cells for all cases is the same.

\begin{figure}[]
    \centering
    \includegraphics[width=0.7\textwidth]{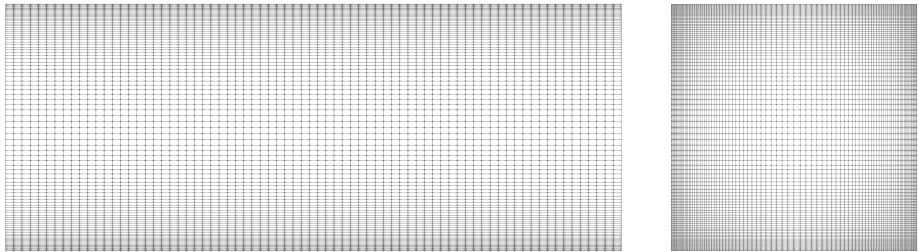}
    \caption{Structured hexahedral mesh used to discretize all square duct cases.}
    \label{fig:duct_mesh}
\end{figure}

The mesh for the square duct case is shown in Figure \ref{fig:duct_mesh}. This mesh is also structured. Cells are concentrated near the boundary layer. The mesh for all square duct cases is identical. The $y^+\leq1$ criterion was verified for the highest Reynolds number flow case. The mesh is 3D, with the dataset for machine learning being generated using a cross-section of the mesh.

\begin{figure}[]
    \centering
    \includegraphics[width=0.7\textwidth]{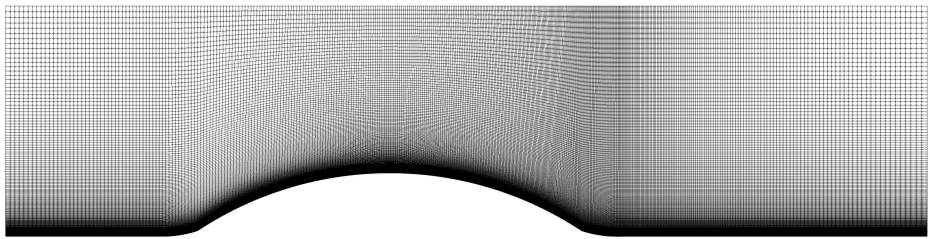}
    \caption{Structured hexahedral mesh used to discretize the $h=42$ mm parametric bump case.}
    \label{fig:bump_mesh}
\end{figure}

\begin{figure}
     \centering
         \includegraphics[width=0.7\textwidth]{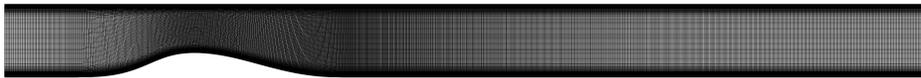}
         \caption{Structured hexahedral mesh for the whole converging-diverging channel case for $Re_H = 20,580$.}\label{fig:cndv_mesh_zoomout}
\end{figure}

\begin{figure}
         \centering
         \includegraphics[width=0.7\textwidth]{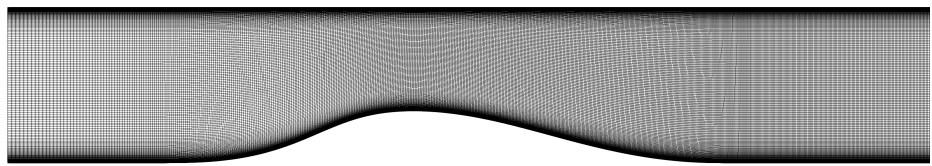}
         \caption{View of the mesh near the region of interest for the converging-diverging channel case for $Re_H = 12,600$. The $Re_H=12,600$ converging-diverging channel case uses a smaller domain than the $Re_H=20,580$ case, but with an identical mesh.}\label{fig:cndv_mesh}
\end{figure}

The parametric bump mesh is shown in Figure \ref{fig:bump_mesh}, and the converging-diverging channel mesh is shown in Figures \ref{fig:cndv_mesh_zoomout} and \ref{fig:cndv_mesh}. Both cases use a structured mesh over an obstruction in the flow. Cells are concentrated in the wake region, and the boundary layer. For the parametric bump, the changing geometry was created by adjusting the bump profile in the structured mesh generator, which resulted in the same number of cells for all cases. The mesh shown in Figure \ref{fig:bump_mesh} is for the highest bump. For the converging-diverging channel, the mesh density for both Reynolds numbers is identical, with the $Re=20,580$ having an extended domain, and therefore more cells.

\begin{figure}[]
    \centering
    \includegraphics[width=0.7\textwidth]{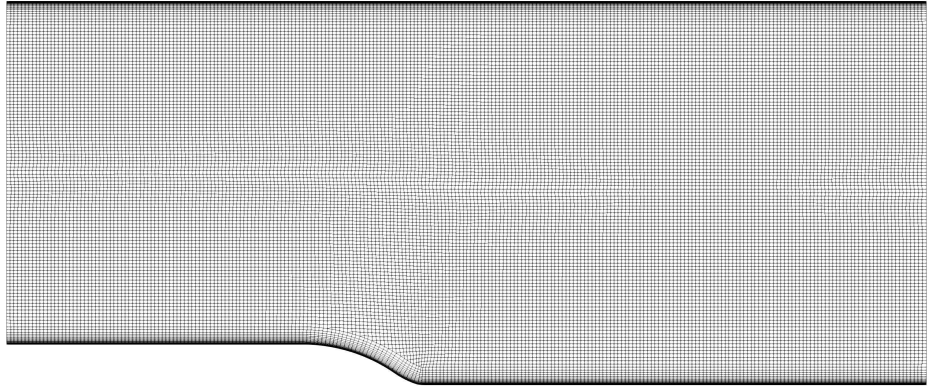}
    \caption{Unstructured, hexahedral dominant mesh used to discretize the curved backward-facing step geometry.}
    \label{fig:cbfs_mesh}
\end{figure}

The only unstructured mesh in the dataset is the curved backward-facing step, shown in Figure \ref{fig:cbfs_mesh}. While it was feasible to generate a structured mesh for this case, an unstructured mesh was generated to include some more typical industrial cells into the dataset. Specifically, near the backward-facing step, the mesh transitions out of the inflation layer using some tetrahedral cells.

\section*{Data Records}



A total of 29 simulations (Table \ref{tbl:cases}) per turbulence model were completed to match the reference data. The DNS or LES reference data were interpolated onto the RANS grid, using linear interpolation. Any points which required extrapolation of the reference data were dropped, and the interpolated reference data were checked for realizability using the criteria from Banerjee et al.~\cite{Banerjee2007}. After interpolation and data quality checks, 895,640 points of RANS data paired with corresponding DNS or LES data are available for each turbulence model. The dataset \cite{McConkey2021} is hosted on Kaggle, a common platform for machine learning.

To maximize the usefulness of the dataset, a comprehensive set of input features and labels was generated. The dataset is organized into two types of data: base variables, and derived quantities provided for convenience. The base variables contain the bare minimum fields that need to be provided to construct the rest of the fields, which are the RANS fields and grid points. The available base fields in the dataset are summarized in Table \ref{tbl:fields_base}, and the derived fields are summarized in Tables \ref{tbl:fields_derived} and \ref{tbl:labels_derived}.

The more useful portion of this dataset is the set of pre-constructed machine learning input features. The selection of input features is a critical area of ongoing research in machine-learnt turbulence models. The typical practice in machine learning Reynolds stress modelling is to derive a set of invariants from a tensor basis, combined with other invariant scalars. This was the approach used in \cite{Ling2016, Wu2018, Kaandorp2018, Xiao2020, Kaandorp2020} and others. While the input feature set varies, an effort has been made to provide sufficient fields in the dataset to conveniently reproduce past feature sets, and develop new ones. For example, all of the input features and labels used by Ling et al.~\cite{Ling2016} are directly provided: the five invariants of the mean strain and rotation rate tensor, the ten basis tensors described in Pope~\cite{Pope1975}, and the anisotropy tensor labels. 

\subsection*{Labels}
This dataset is suited for models that predict the Reynolds stress tensor, an equivalent problem to predicting the anisotropy tensor. The provided label set includes the individual Reynolds stress components (the base labels), and other fields that are sometimes more convenient to use. The Reynolds stress tensor, TKE, and anisotropy tenor are provided as ready-to-use labels. 

\subsection*{Invariants of tensor bases}
The invariants are derived from a set of basis tensors, which form a basis for the space spanned by a set of feature tensors. First, the feature tensors need to be selected. The selection of the feature tensors determines what flow variable gradients are incorporated into the model. Previous investigations have selected the set of feature tensors as $\{\hat{S},\hat{R}\}$ \cite{Ling2016}, $\{\hat{S},\hat{R},\nabla k\}$ \cite{Kaandorp2018, Kaandorp2020}, and $\{\hat{S},\hat{R},\nabla k,\nabla p\}$ \cite{Wu2018}. If the feature tensors were directly employed as input features, the model would not be invariant because these inputs change with the coordinate system. Therefore, the procedure presented by Spencer and Rivlin~\cite{Spencer1962} is commonly employed to generate a tensor basis for the feature set. After constructing the tensor basis, the invariants of the tensor basis are taken --- in other words, the traces of the basis tensors are used as input features. This procedure guarantees that the model has the same invariance properties as the trace of the basis tensors.

The dataset includes several quantities which are convenient in generating tensor bases. Along with the velocity gradient tensor $\nabla U$, the strain rate and rotation rate tensors $S,R$ are provided. While the strain and rotation rate tensors are provided without normalization, a set of pre-normalized strain and rotation rate tensors $\hat{S},\hat{R}$ are provided, with the normalizations shown in Table \ref{tbl:fields_derived}. A similar set of features for the kinematic pressure and TKE gradients are provided. The gradients themselves, a vector quantity, and the associated antisymmetric tensors for both the un-normalized and normalized forms are provided.

The provided dataset is sufficient to form the most comprehensive tensor bases used to date, which is the 47 tensor basis used by Wu et al.~\cite{Wu2018}. However, it is the traces of these 47 tensors which are of interest. These 47 invariant traces are included in the dataset to be directly used as input features to a machine learning model. Also included is the set of 5 invariants ($\lambda_i$), which arise from using the strain and rotation rate as the feature tensors, as in Ling et al.~\cite{Ling2016}.

\subsection*{Other input scalars}

After gathering the set of tensor basis invariants, an additional set of scalars is added. Care must be taken that these scalars are invariant to not corrupt the invariance of the constructed tensor basis invariants. While many scalars have been proposed, many of them are not Galilean invariant, which is a property desired in machine-learnt turbulence models. Therefore, four Galilean invariant scalars used by Kaandorp and Dwight \cite{Kaandorp2020} are included as ready-to-use features in the dataset. While this set of input scalars is not comprehensive, the dataset includes sufficient fields to conveniently generate other scalar quantities.

\section*{Technical Validation}


Iterative residual convergence below $10^{-6}$ was generally achieved, with most simulations converging below $10^{-8}$. The residual plots for each simulation are provided along with the dataset. The exceptions to this tight residual convergence criteria are the $U_y$ $U_z$, and $p$ fields for the square duct cases. The linear eddy viscosity model is unable to accurately predict the secondary vortices resulting from non-zero $U_y$ and $U_z$ components in the square duct case, and therefore minimal convergence is seen in these residuals as the in-plane velocity fields remain close to the initial condition of zero. The pressure field for the square duct case does not converge below $10^{-6}$ due to the presence of a forcing term which maintains the bulk velocity, resulting in uniform streamwise zero pressure equal to the initial condition of zero.

The RANS results are sensitive to the mesh used. While the mesh must be compatible with the selected wall treatment, it must also be sufficiently fine to reduce discretization errors. To demonstrate that the selected meshes do not affect the result, a mesh independence study was completed for each of the five flow cases. The most demanding case was selected for each flow type: the steepest periodic hills case, the highest Reynolds number square duct, the highest bump, the highest Reynolds number converging-diverging channel, and the curved backward-facing step. Mesh independence was demonstrated using the $k$-$\varepsilon$ turbulence model. The mesh study was conducted by examining the change in the velocity fields between varying mesh sizes. 

\begin{figure}[]
    \centering
    \includegraphics[]{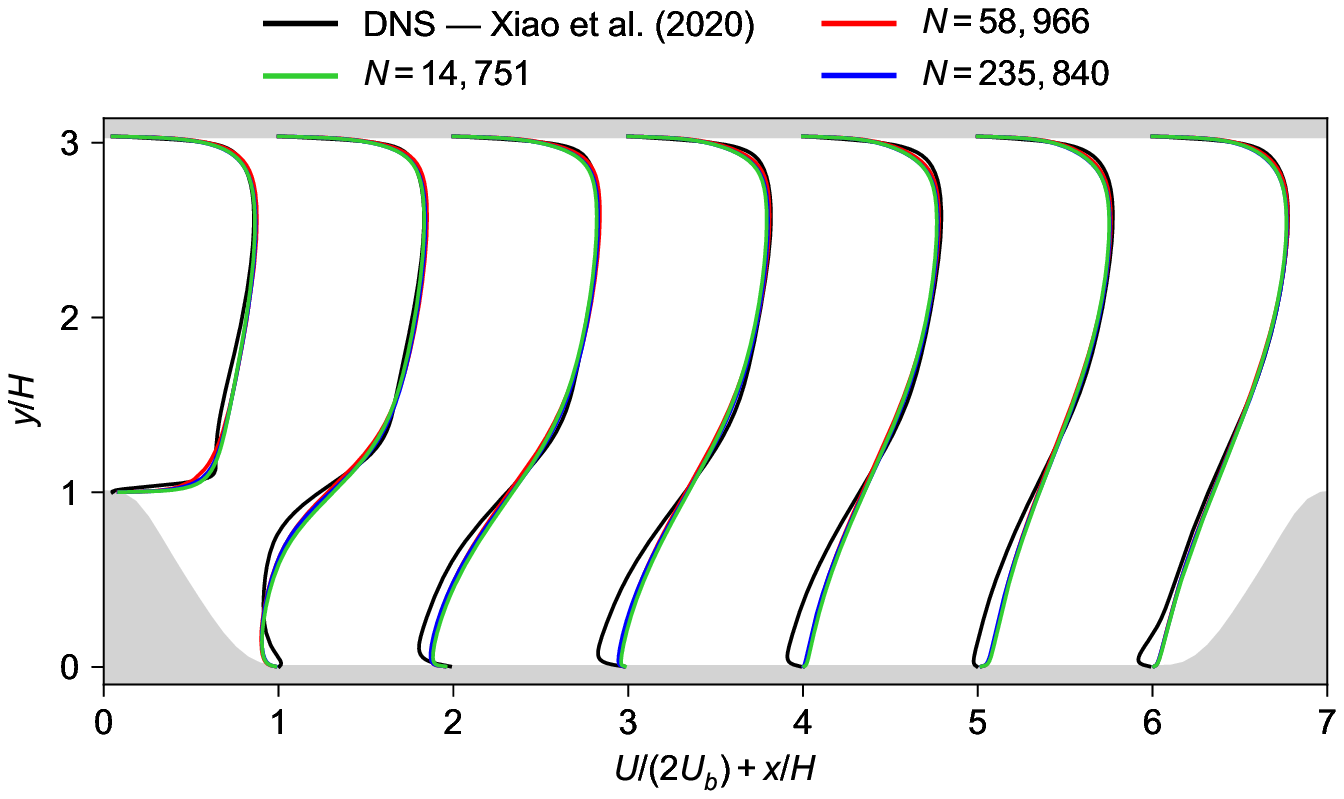}
    \caption{Profiles of $U$ for three meshes of varying density for the $\alpha=0.5$ periodic hills case.}
    \label{fig:phll_mesh_conv_U}
\end{figure}

\begin{figure}[]
    \centering
    \includegraphics[]{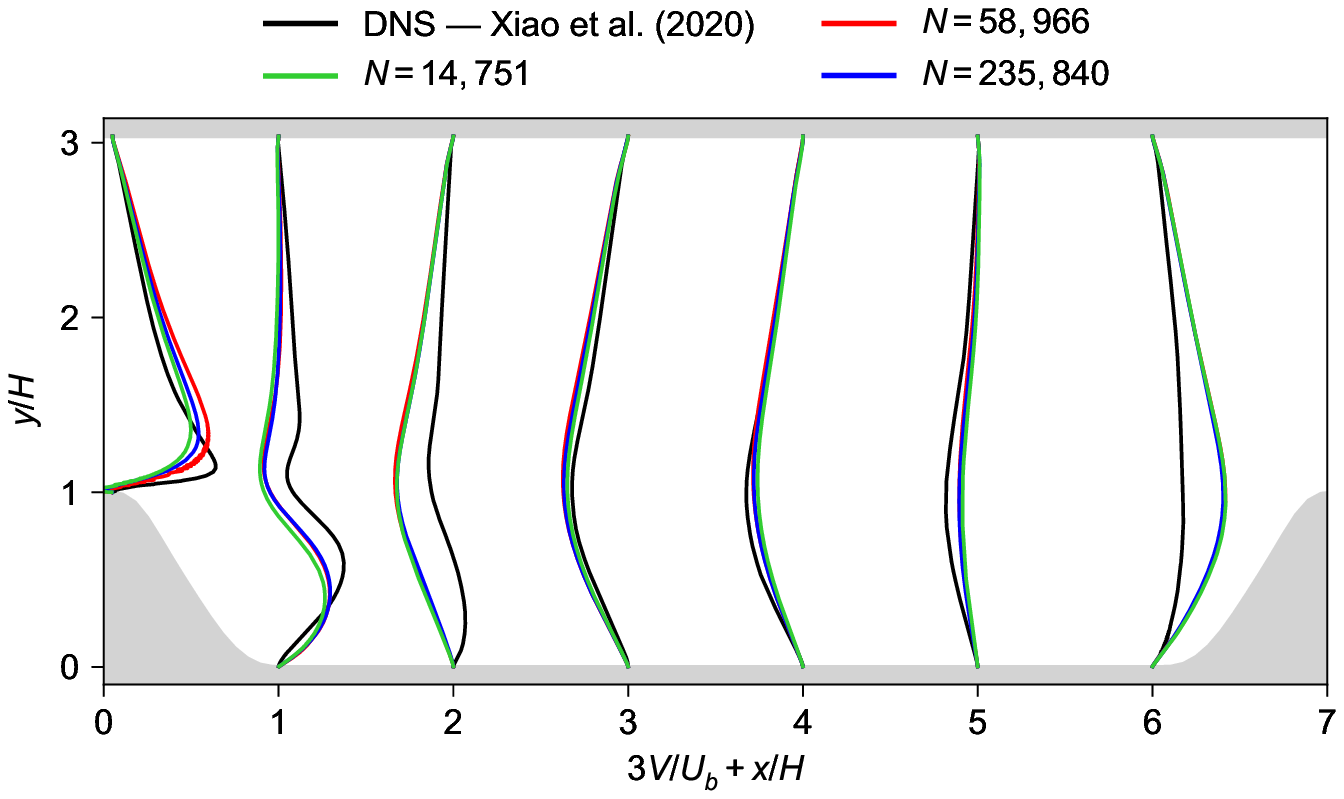}
    \caption{Profiles of $V$ for three meshes of varying density for the $\alpha=0.5$ periodic hills case.}
    \label{fig:phll_mesh_conv_V}
\end{figure}

Figures \ref{fig:phll_mesh_conv_U} and \ref{fig:phll_mesh_conv_V} show the results of the mesh convergence study for the periodic hills case. The meshes provided by Xiao et al.~\cite{Xiao2020} were refined two times, each by a factor of 2 in the $x$ and $y$ directions. A small group of cells could not be refined while maintaining reasonable quality, which is why the meshes shown in Figures \ref{fig:phll_mesh_conv_U} and \ref{fig:phll_mesh_conv_V} do not exactly contain $N,4N,$ and $16N$ cells. The results for the periodic hills case demonstrate good mesh convergence for the grid with the smallest number of cells used in the study. There is almost no change for the $U$ velocity for grids whose number of cells is greater than $N=14,751$. The $V$ profiles near the inlet boundary shown changes between the mesh sizes. For this case, the mesh convergence is non-monotonic, but the differences of the $V$ profiles between the various meshes used are small. Therefore, the $N=14,751$ mesh is sufficiently converged.

\begin{figure}[]
    \centering
    \includegraphics[]{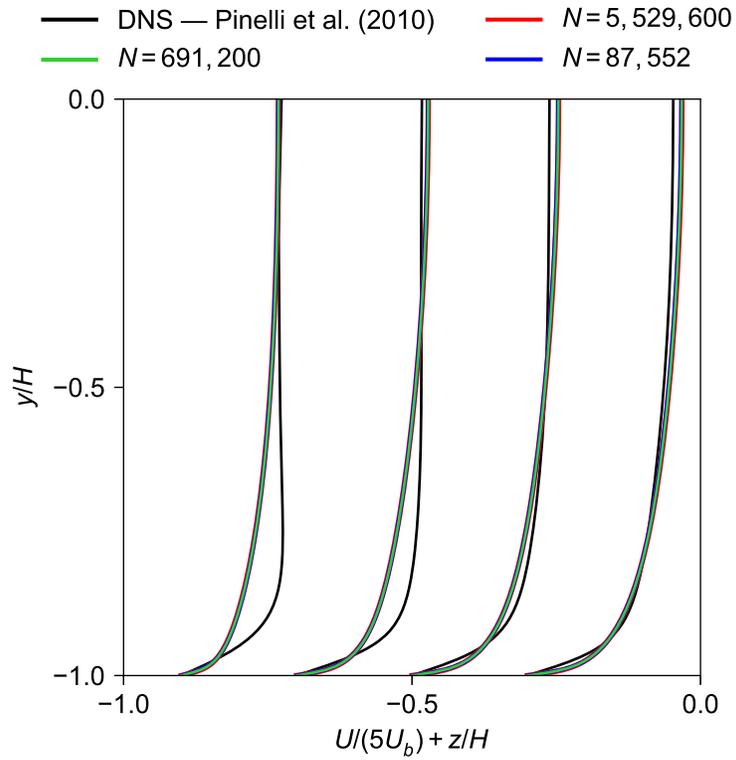}
    \caption{Profiles of $U$ for three meshes of varying density for the $Re=3,500$ square duct case.}
    \label{fig:duct_mesh_conv_U}
\end{figure}

One of the main considerations for the square duct mesh is sufficient resolution in the $y-z$ plane to extract machine learning features. The reference data by Pinelli et al.~\cite{Pinelli2010} are provided as a set of statistics in the $y-z$ plane. Even though Figure \ref{fig:duct_mesh_conv_U} shows that the solution is mesh-converged at $N=87,552$, the resolution in the $y-z$ plane is too coarse. The $N=87,552$ mesh results in 2,304 dataset points per case, while the $N=691,200$ mesh results in 9,216 points per case. Therefore, the $N=691,200$ mesh is selected for generating the dataset, because the solution is mesh independent, and there are sufficient cells in the $y-z$ plane to generate features for machine learning.

\begin{figure}[]
    \centering
    \includegraphics[]{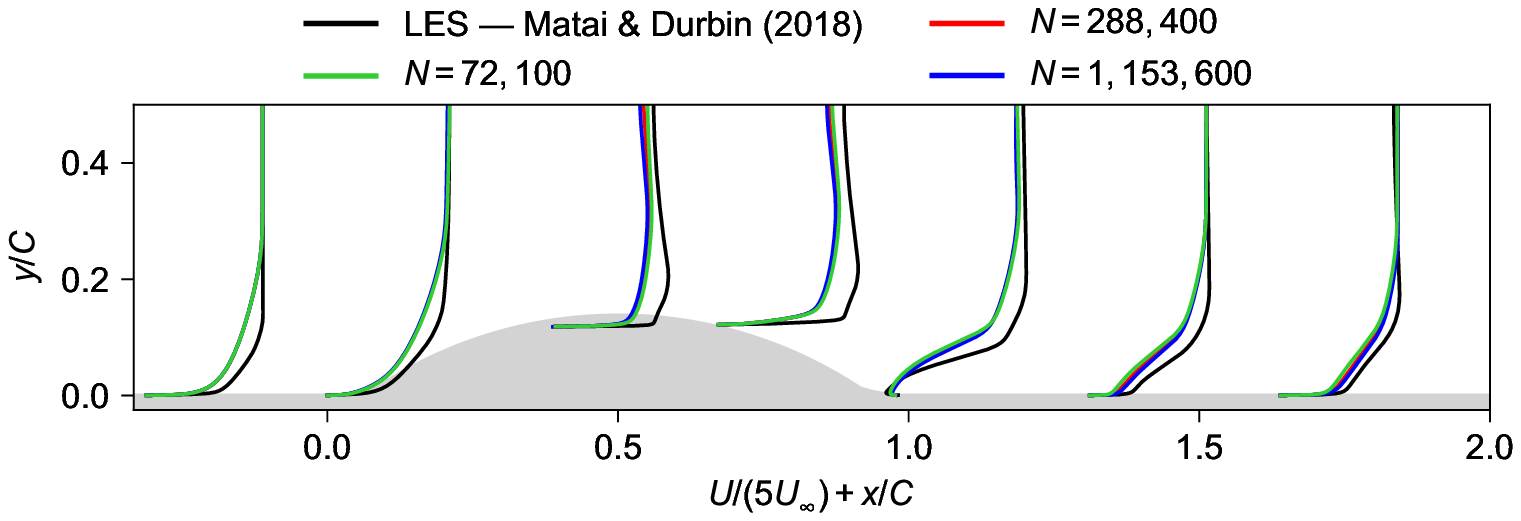}
    \caption{Profiles of $U$ for three meshes of varying density for the $h=42$ mm parametric bump case.}
    \label{fig:bump_mesh_conv_U}
\end{figure}

\begin{figure}[]
    \centering
    \includegraphics[]{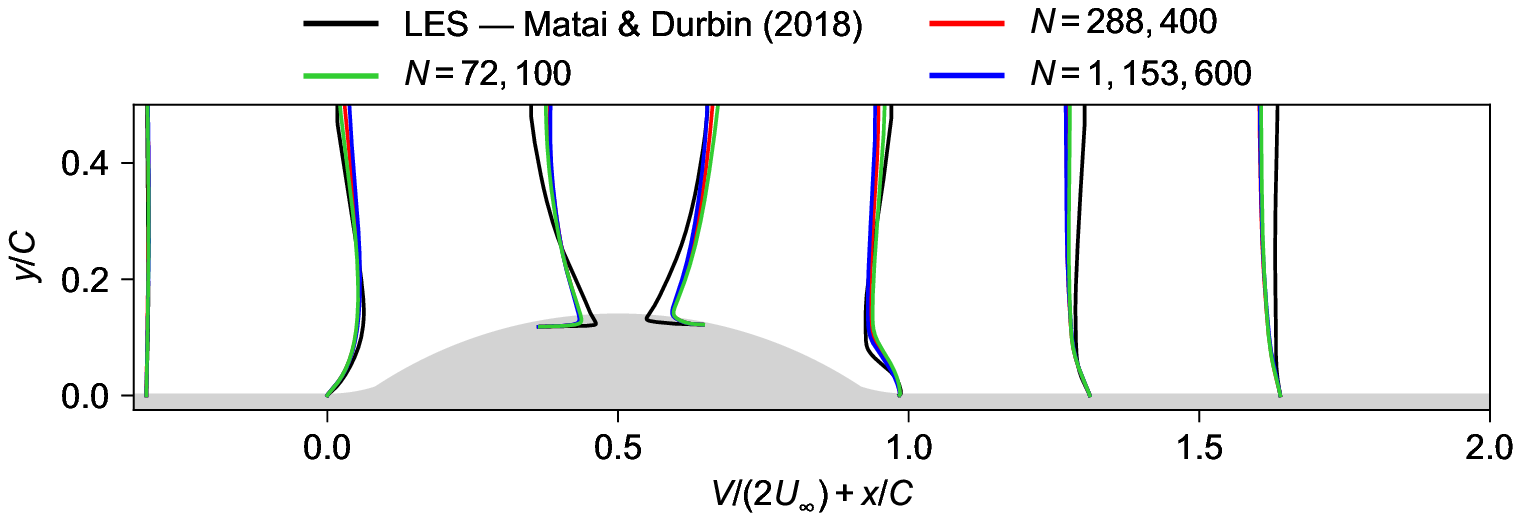}
    \caption{Profiles of $U$ for three meshes of varying density for the $h=42$ mm parametric bump case.}
    \label{fig:bump_mesh_conv_V}
\end{figure}

The parametric bump is the highest Reynolds number flow in the dataset ($Re_H\approx27,850$) and, as a consequence, it requires a dense mesh. Solution convergence at the coarsest mesh with $N = 72,100$ cells was demonstrated by increasing the number of cells in the structured mesh generator by a factor of two, and then four, and comparing the velocity profiles for the corresponding $N,4N,$ and $16N$ cases. Figures \ref{fig:bump_mesh_conv_U} and \ref{fig:bump_mesh_conv_V} show the comparisons made. For the $U$ velocity profile, there are small differences in the wake of the bump, and in the far-field above the bump. The $V$ velocity field reflects these small far-field differences above the bump. However, the differences are comparatively small, and the mesh demonstrates good convergence to generate the dataset.

\begin{figure}[]
    \centering
    \includegraphics[]{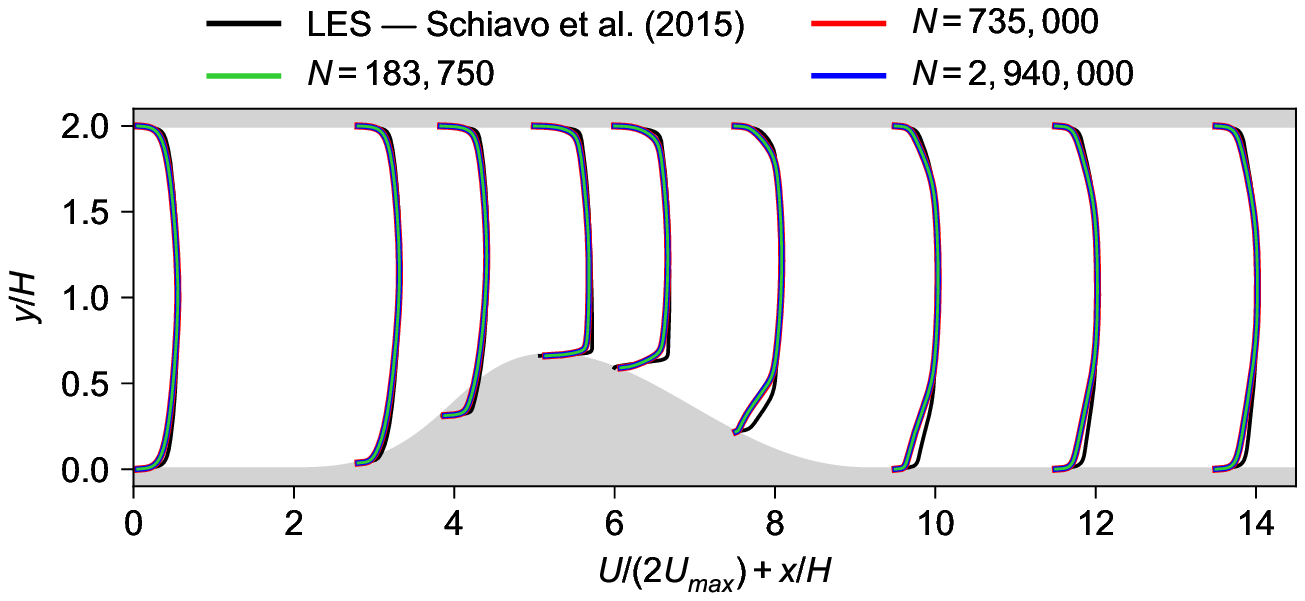}
    \caption{Profiles of $U$ for three meshes of varying density for the $Re=20,580$ converging-diverging channel case.}
    \label{fig:cndv_mesh_conv_U}
\end{figure}

\begin{figure}[]
    \centering
    \includegraphics[]{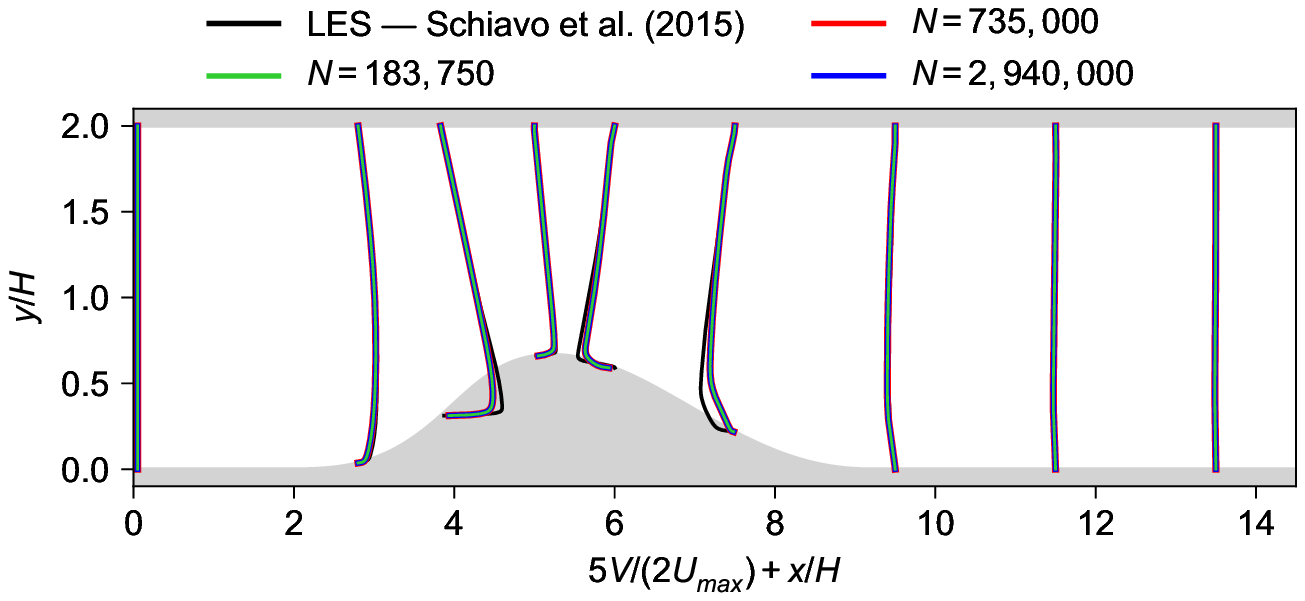}
    \caption{Profiles of $V$ for three meshes of varying density for the $Re=20,580$ converging-diverging channel case.}
    \label{fig:cndv_mesh_conv_V}
\end{figure}

Mesh convergence for the converging-diverging channel case was demonstrated similarly to the bump case. The number of cells in the structured mesh generator was increased by a factor of two, and then four. Figures \ref{fig:cndv_mesh_conv_U} and \ref{fig:cndv_mesh_conv_V} show that there are almost no differences between the solutions as the mesh is refined, even by a factor of 16. Therefore, the mesh for the converging-diverging channel case is sufficiently converged at $N=183,750$.

\begin{figure}[]
    \centering
    \includegraphics[]{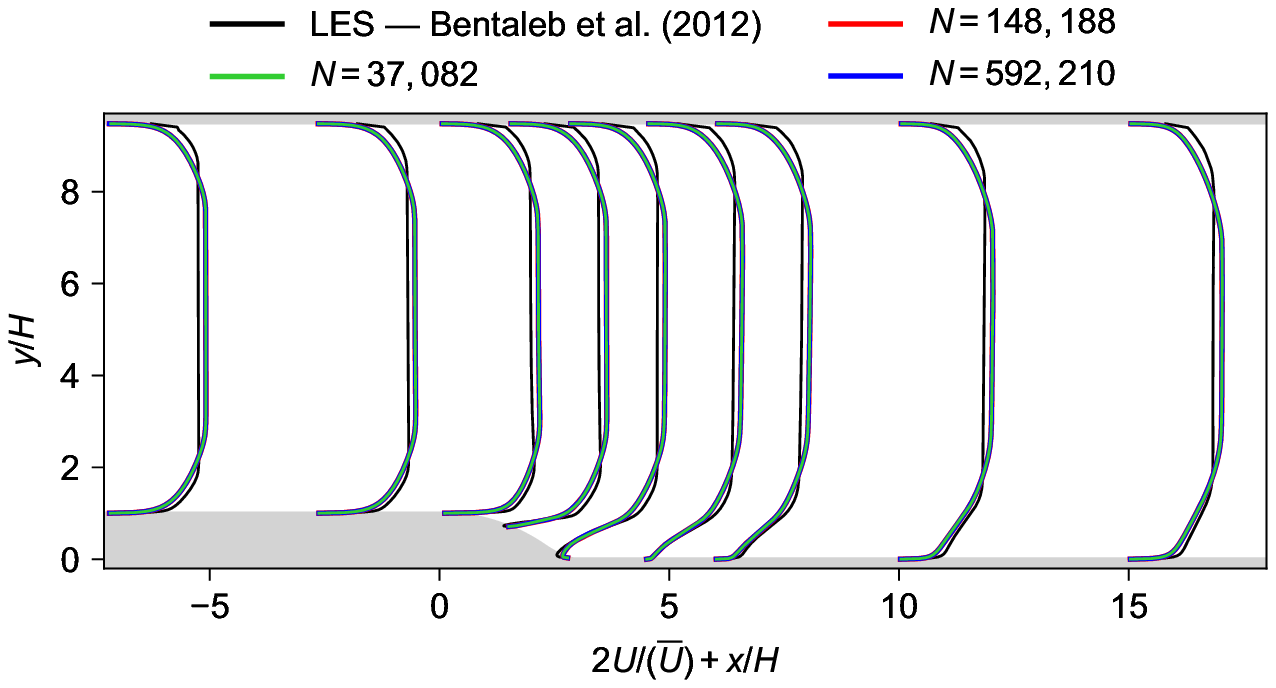}
    \caption{Profiles of $U$ for three meshes of varying density for the curved backward-facing step case.}
    \label{fig:cbfs_mesh_conv_U}
\end{figure}

\begin{figure}[]
    \centering
    \includegraphics[]{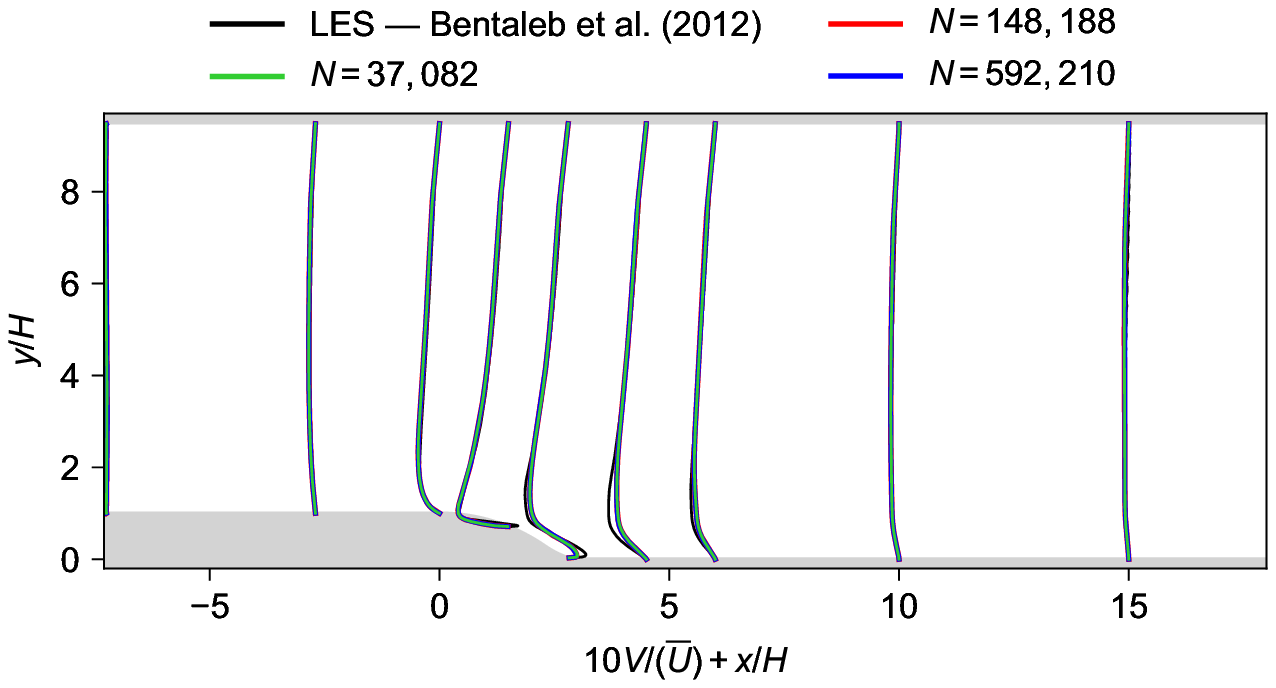}
    \caption{Profiles of $V$ for three meshes of varying density for the curved backward-facing step case.}
    \label{fig:cbfs_mesh_conv_V}
\end{figure}

The curved backward-facing step case utilizes an unstructured mesh, with a small number of tetrahedral cells. Demonstrating mesh convergence was completed similarly to the periodic hills case, by refining the mesh twice in each direction. Some cells could not be refined while maintaining reasonable mesh quality, which is the reason that the meshes in Figures \ref{fig:cbfs_mesh_conv_U} and \ref{fig:cbfs_mesh_conv_V} do not exactly have $N,\ 4N,$ and $16N$ cells. The solution has excellent mesh convergence at $N=37,082$, in both the $U$ and $V$ velocity fields.

\section*{Usage Notes}



The dataset structure consists of a folder for each turbulence model, with an additional folder for the DNS/LES labels \cite{McConkey2021}. The RANS features for each case are provided using a consistent naming scheme. This structure allows the data to be accessed and processed in a coherent manner for immediate use in open-source machine learning frameworks such as TensorFlow and PyTorch. An example of how to use the data to develop a simple machine learning model for the Reynolds stress anisotropy tensor are provided on the dataset page. The dataset will be updated as more DNS/LES reference datasets become available, or if there is demand to include additional RANS turbulence models. 

There are approximately 1,000 fields per turbulence model, provided as \texttt{numpy} arrays. The first index for all fields in the dataset is the data point index, equivalent to the cell index. The remaining indices in the array depends on the nature of the field. For example, all tensors are given with shape $(N,3,3)$, where $N$ is the data point index. The ten basis tensors used in a general representation of the anisotropy tensor proposed by Pope \cite{Pope1975} are given as an array with shape $(N,10,3,3)$. Relatively few pre-processing steps have been performed on the dataset --- no normalization or outlier elimination has been performed. The only deletions arise from a small subset (less than 50 points) of non-realizable LES label values, and any points requiring extrapolation of the reference data. Therefore, it is recommended that after a specific input feature set is formed using the provided fields, the input features should be standardized as is typical in machine learning. The RANS results also contain some outliers that may need to be dropped. For example, Kaandorp \cite{Kaandorp2018} dropped datapoints outside of $\mu \pm 5\sigma$, where $\mu$ is the mean, and $\sigma$ is the standard deviation.

\section*{Code availability}

Both the code used for generating this dataset and input files for the OpenFOAM simulations are available on the Kaggle page for this dataset \cite{McConkey2021}. The software used was OpenFOAM v2006, with all scripts written in Python 3. 

\bibliography{references}


\section*{Acknowledgements} 

R.M is supported by the Ontario Graduate Scholarship program (OGS). The computational resources for this work were supported by the Tyler Lewis Clean Energy Research Foundation (TLCERF). 

\section*{Author contributions statement}


All authors conceived the experiments. R.M conducted the simulations, and prepared the dataset. All authors reviewed the manuscript.

\section*{Competing interests} 

The authors declare no competing interests.

\clearpage
\section*{Figures \& Tables}




\begin{table}[]\caption{Inputs and outputs of the present study.}\label{tbl:inputsoutputs}
\centering
\begin{tabular}{ccc}
\hline
\multicolumn{2}{c}{Inputs} &
  Outputs \\ \hline
Present work &
  \multicolumn{1}{c|}{Previous work} &
  \multirow{2}{*}{\begin{tabular}[c]{@{}c@{}}A set of \textbf{features} and \textbf{labels} for developing models\\ which map the coarse variables to highly-resolved variables\\ \textbf{Features}: Coarsely-resolved flow fields\\ with curated machine learning input features\\ \textbf{Labels}: Highly-resolved flow fields, mapped onto\\ the coarse grid, with curated machine learning labels\end{tabular}} \\ \cline{1-2}
\begin{tabular}[c]{@{}c@{}}Numerical settings \\ (e.g. schemes, grids) \\ for generating \\ coarsely-resolved\\  flow fields\end{tabular} &
  \multicolumn{1}{c|}{\begin{tabular}[c]{@{}c@{}}Highly-resolved \\ flow fields,\\ suitable for use \\ as "truth" values\\ in machine learning\end{tabular}} &
   \\ \hline
\end{tabular}
\end{table}

\begin{table}[]\caption{Cases in the dataset. $Re_L$ is the Reynolds number based on the characteristic length and velocity scales shown in Figures \ref{fig:periodichills} to \ref{fig:cbfs}.}\label{tbl:cases}
\centering
\begin{tabular}{cccccc}
\hline
Flow case                                                              & Ref.         & $Re_L$        & Dim. & Num. cases & Parameter   \\ \hline
Periodic hills                                                         & \cite{Xiao2020} & $5,600$          & 2D   & 5          & Steepness   \\
Square duct                                                            & \cite{Pinelli2010}         & 1,100--3,500     & 3D   & 16         & $Re$        \\
Parametric bump                                                        & \cite{Matai2019a}        & 13,260--27,850 & 2D   & 5          & Bump height \\
\begin{tabular}[c]{@{}c@{}}Converging-diverging\\ channel\end{tabular} & \cite{Laval2011, Schiavo2015}         & 12,600--20,580 & 2D   & 2          & $Re$        \\
\begin{tabular}[c]{@{}c@{}}Curved backward-facing \\ step\end{tabular} & \cite{Bentaleb2012}         & 13,700        & 2D   & 1          & -           \\ \hline
\end{tabular}
\end{table}
\begin{table}[]\caption{Kinematic molecular viscosity used for each case.}\label{tbl:viscosity}
\centering
\begin{tabular}{cc}
\hline
Flow case                                                              & $\nu$ (m$^2$/s)         \\ \hline
Periodic hills                                                         &   $5(10)^{-6}$ \\
Square duct                                                            &   0.241/$Re_H$    \\
Parametric bump                                                        &  $2.529(10)^{-5}$\\
Converging-diverging channel & $1/Re_H$  \\
Curved backward-facing step &     $7.3(10)^{-5}$  \\ \hline
\end{tabular}
\end{table}
\begin{table}[]
\centering
\caption{Units for each variable requiring boundary conditions.}
\label{tbl:units}
\begin{tabular}{ccc}
\hline
Description &Field             & Units    \\ \hline
Velocity &$\vec{U}$           &   m/s        \\
Kinematic pressure &$p$           &       m$^{2}$/s$^{2}$             \\
Turbulent kinetic energy&$k$           &  m$^{2}$/s$^{2}$\\
TKE dissipation rate&$\varepsilon$ &  m$^{2}$/s$^{3}$\\
TKE specific dissipation rate&$\omega$ &   s$^{-1}$                           \\
Anisotropy measure& $\phi_t$      & -                                   \\
TKE redistribution scalar&$f$  & s$^{-1}$                                                  \\ \hline
\end{tabular}
\end{table}
\begin{table}[]
\centering
\caption{Boundary conditions for the periodic hills and square duct cases.}
\label{tbl:duct}
\begin{tabular}{ccccc}
\hline
              & Inlet  & Outlet & Walls                                    \\ \hline
$\vec{U}$           & Cyclic & Cyclic & $\vec{U}=0$                                    \\
$p$           & Cyclic & Cyclic & Zero-gradient                              \\
$k$           & Cyclic & Cyclic & $k=0$             & \\
$\varepsilon$ & Cyclic & Cyclic & $\varepsilon=\varepsilon_{vis} = 2wk\nu/y^2$ \\
$\omega$      & Cyclic & Cyclic & $\omega= 6\nu/(\beta_1 y^2)$                         \\
$\phi_t$      & Cyclic & Cyclic & $\phi_t$ = 0                                             \\
$f$           & Cyclic & Cyclic & $f$ = 0                                             \\ \hline
\end{tabular}
\end{table}
\begin{table}[]
\centering
\caption{Inlet conditions for the flat developing flow case, used to generate an inlet profile for the bump cases.}
\label{tbl:bump_develop}
\begin{tabular}{ccccc}
\hline
              & Inlet                            & Outlet & Top & Bottom\\ \hline
$\vec{U}$           & $\vec{U}$ = (16.683, 0, 0)      & Zero-gradient & Zero-gradient &$\vec{U}=0$  \\
$p$           & Zero-gradient                     & $p=0$ & Zero-gradient &Zero-gradient\\
$k$           & $k = 0.16699 (I=2\%)$  & Zero-gradient & Zero-gradient &$k=0$ \\
$\varepsilon$ & $\varepsilon = 0.266974$ ($L_t=42$ mm) & Zero-gradient & Zero-gradient&$\varepsilon=\varepsilon_{vis} = 2wk\nu/y^2$\\
$\omega$      & $\omega= 17.764$             & Zero-gradient & Zero-gradient&$\omega= 6\nu/(\beta_1 y^2)$\\
$\phi_t$      & Zero-gradient                     & Zero-gradient & Zero-gradient&$\phi_t$ = 0 \\
$f$           & Zero-gradient                       & Zero-gradient & Zero-gradient&$f$ = 0 \\ \hline
\end{tabular}
\end{table}

\begin{table}[]
\centering
\caption{Boundary conditions for the bump cases.}
\label{tbl:bump}
\begin{tabular}{ccccc}
\hline
              & Inlet                            & Outlet & Top & Bottom\\ \hline
$\vec{U}$           & Fully-developed, $U_\infty=16.77$ m/s      & Zero-gradient & Zero-gradient &$\vec{U}=0$  \\
$p$           & Zero-gradient                     & $p=0$ & Zero-gradient &Zero-gradient\\
$k$           & Fully-developed  & Zero-gradient & Zero-gradient &$k=0$ \\
$\varepsilon$ & Fully-developed & Zero-gradient & Zero-gradient&$\varepsilon=\varepsilon_{vis} = 2wk\nu/y^2$\\
$\omega$      & Fully-developed             & Zero-gradient & Zero-gradient&$\omega= 6\nu/(\beta_1 y^2)$\\
$\phi_t$      & Zero-gradient                     & Zero-gradient & Zero-gradient&$\phi_t$ = 0 \\
$f$           & Zero-gradient                       & Zero-gradient & Zero-gradient&$f$ = 0 \\ \hline
\end{tabular}
\end{table}
\begin{table}[]
\centering
\caption{Inlet conditions for the flat developing flow case, used to generate an inlet profile for the converging-diverging channel cases.}
\label{tbl:cndv_develop}
\begin{tabular}{ccccc}
\hline
              & Inlet                            & Outlet & Top & Bottom\\ \hline
$\vec{U}$           & $\vec{U}$ = (0.845, 0, 0)      & Zero-gradient & $\vec{U}=0$ &$\vec{U}=0$  \\
$p$           & Zero-gradient                     & $p=0$ & Zero-gradient &Zero-gradient\\
$k$           & $k = 4.28421(10)^{-4} (I=2\%)$  & Zero-gradient & $k=0$ &$k=0$ \\
$\varepsilon$ & $\varepsilon = 1.0408(10)^{-5}$ ($L_t=0.07H_\text{chan}$) & Zero-gradient & $\varepsilon=\varepsilon_{vis} = 2wk\nu/y^2$&$\varepsilon=\varepsilon_{vis} = 2wk\nu/y^2$\\
$\omega$      & $\omega= 0.26993$             & Zero-gradient & $\omega= 6\nu/(\beta_1 y^2)$&$\omega= 6\nu/(\beta_1 y^2)$\\
$\phi_t$      & Zero-gradient                     & Zero-gradient & $\phi_t$ = 0&$\phi_t$ = 0 \\
$f$           & Zero-gradient                       & Zero-gradient & $f$ = 0&$f$ = 0 \\ \hline
\end{tabular}
\end{table}

\begin{table}[]
\centering
\caption{Boundary conditions for the converging-diverging channel cases.}
\label{tbl:cndv}
\begin{tabular}{ccccc}
\hline
              & Inlet                            & Outlet & Top & Bottom\\ \hline
$\vec{U}$           & Fully-developed, $U_\text{max}=1.0$ m/s      & Zero-gradient & $\vec{U}=0$ &$\vec{U}=0$  \\
$p$           & Zero-gradient                     & $p=0$ & Zero-gradient &Zero-gradient\\
$k$           & Fully-developed  & Zero-gradient & $k=0$ &$k=0$ \\
$\varepsilon$ & Fully-developed & Zero-gradient & $\varepsilon=\varepsilon_{vis} = 2wk\nu/y^2$&$\varepsilon=\varepsilon_{vis} = 2wk\nu/y^2$\\
$\omega$      & Fully-developed             & Zero-gradient & $\omega= 6\nu/(\beta_1 y^2)$&$\omega= 6\nu/(\beta_1 y^2)$\\
$\phi_t$      & Zero-gradient                     & Zero-gradient & $\phi_t$ = 0&$\phi_t$ = 0 \\
$f$           & Zero-gradient                       & Zero-gradient & $f$ = 0&$f$ = 0 \\ \hline
\end{tabular}
\end{table}

\begin{table}[]
\centering
\caption{Inlet conditions for the flat developing flow case, used to generate an inlet profile for the curved backward facing step cases.}
\label{tbl:cbfs_develop}
\begin{tabular}{ccccc}
\hline
              & Inlet                            & Outlet & Top & Bottom\\ \hline
$\vec{U}$           & $\vec{U}$ = (1.0, 0, 0)      & Zero-gradient & $\vec{U}=0$ &$\vec{U}=0$  \\
$p$           & Zero-gradient                     & $p=0$ & Zero-gradient &Zero-gradient\\
$k$           & $k = 6.00(10)^{-4} (I=2\%)$  & Zero-gradient & $k=0$ &$k=0$ \\
$\varepsilon$ & $\varepsilon = 2.415(10)^{-6}$ ($L_t=H$) & Zero-gradient & $\varepsilon=\varepsilon_{vis} = 2wk\nu/y^2$&$\varepsilon=\varepsilon_{vis} = 2wk\nu/y^2$\\
$\omega$      & $\omega= 4.472(10)^{-2}$             & Zero-gradient & $\omega= 6\nu/(\beta_1 y^2)$&$\omega= 6\nu/(\beta_1 y^2)$\\
$\phi_t$      & Zero-gradient                     & Zero-gradient & $\phi_t$ = 0&$\phi_t$ = 0 \\
$f$           & Zero-gradient                       & Zero-gradient & $f$ = 0&$f$ = 0 \\ \hline
\end{tabular}
\end{table}
\begin{table}[]
\centering
\caption{Boundary conditions for curved backward facing step case.}
\label{tbl:cbfs}
\begin{tabular}{ccccc}
\hline
              & Inlet                            & Outlet & Top & Bottom\\ \hline
$\vec{U}$           & Fully-developed, $\overline{U}=1.0$ m/s      & Zero-gradient & $\vec{U}=0$ &$\vec{U}=0$  \\
$p$           & Zero-gradient                     & $p=0$ & Zero-gradient &Zero-gradient\\
$k$           & Fully-developed  & Zero-gradient & $k=0$ &$k=0$ \\
$\varepsilon$ & Fully-developed & Zero-gradient & $\varepsilon=\varepsilon_{vis} = 2wk\nu/y^2$&$\varepsilon=\varepsilon_{vis} = 2wk\nu/y^2$\\
$\omega$      & Fully-developed             & Zero-gradient & $\omega= 6\nu/(\beta_1 y^2)$&$\omega= 6\nu/(\beta_1 y^2)$\\
$\phi_t$      & Zero-gradient                     & Zero-gradient & $\phi_t$ = 0&$\phi_t$ = 0 \\
$f$           & Zero-gradient                       & Zero-gradient & $f$ = 0&$f$ = 0 \\ \hline
\end{tabular}
\end{table}
\begin{table}[]\caption{Meshes used for discretizing the domain.}\label{tbl:mesh}
\centering
\begin{tabular}{ccccc}
\hline
Case                                                                   & Dim. & Mesh type                                                                   & $N$     & Generation method                                 \\ \hline
Periodic hills                                                         & 2D   & \begin{tabular}[c]{@{}c@{}}Structured\\ hexahedral\end{tabular}             & 14,751  & Provided by Xiao et al.~\cite{Xiao2020} \\
Square duct                                                            & 3D   & \begin{tabular}[c]{@{}c@{}}Structured\\ hexahedral\end{tabular}             & 691,300 & \texttt{blockMesh}\cite{Openfoam}               \\
Parametric bump                                                        & 2D   & \begin{tabular}[c]{@{}c@{}}Structured\\ hexahedral\end{tabular}             & 72,100  & \texttt{blockMesh}\cite{Openfoam}               \\
\begin{tabular}[c]{@{}c@{}}Converging-diverging\\ channel\end{tabular} & 2D   & \begin{tabular}[c]{@{}c@{}}Structured\\ hexahedral\end{tabular}             & 183,750 & \texttt{blockMesh}\cite{Openfoam}               \\
\begin{tabular}[c]{@{}c@{}}Curved backward-facing \\ step\end{tabular} & 2D   & \begin{tabular}[c]{@{}c@{}}Unstructured \\ hexahedral dominant\end{tabular} & 37,082  & \texttt{snappyHexMesh}\cite{Openfoam}          \\ \hline
\end{tabular}
\end{table}

\begin{table}[]\caption{Base fields available in the dataset.}\label{tbl:fields_base}
\centering
\begin{tabular}{cccc}
\hline
Quantity                                                                                                     & Units       & Symbol            & Fieldname        \\ \hline
\multicolumn{4}{c}{Features from RANS}                                                                                                                            \\ \hline
$x$-coordinate                                                                                               & m           & $x$               & \texttt{x}       \\
$y$-coordinate                                                                                               & m           & $y$               & \texttt{y}       \\
$z$-coordinate                                                                                               & m           & $z$               & \texttt{z}       \\
$x$ velocity component                                                                                       & m/s         & $U_x$             & \texttt{Ux}      \\
$y$ velocity component                                                                                       & m/s         & $U_y$             & \texttt{Uy}      \\
$z$ velocity component                                                                                       & m/s         & $U_z$             & \texttt{Uz}      \\
Kinematic pressure                                                                                           & m$^2$/s$^2$         & $p$               & \texttt{p}       \\
Turbulent kinetic energy                                                                                     & m$^2$/s$^2$ & $k$               & \texttt{k}       \\
TKE dissipation rate                                                                                         & m$^2$/s$^3$ & $\varepsilon$     & \texttt{epsilon} \\
TKE specific dissipation rate                                                                                & s$^{-1}$    & $\omega$          & \texttt{omega}   \\
Anisotropy measure & -           & $\phi_t$         & \texttt{phit}    \\
TKE redistribution scalar                                                                                     & s$^{-1}$    & $f$               & \texttt{f}       \\ \hline
\multicolumn{4}{c}{Labels from DNS/LES}                                                                                                                           \\ \hline
$x$ mean velocity component                                                                                  & m/s         & $\overline{u}$    & \texttt{um}      \\
$y$ mean velocity component                                                                                  & m/s         & $\overline{v}$    & \texttt{vm}      \\
$z$ mean velocity component                                                                                  & m/s         & $\overline{w}$    & \texttt{wm}      \\
$x$ Reynolds normal stress                                                                                   & m$^2$/s$^2$ & $\overline{u'u'}$ & \texttt{uu}      \\
$xy$ Reynolds shear stress                                                                                   & m$^2$/s$^2$ & $\overline{u'v'}$ & \texttt{uv}      \\
$xz$ Reynolds shear stress                                                                                   & m$^2$/s$^2$ & $\overline{u'w'}$ & \texttt{uw}      \\
$y$ Reynolds normal stress                                                                                   & m$^2$/s$^2$ & $\overline{v'v'}$ & \texttt{vv}      \\
$yz$ Reynolds shear stress                                                                                   & m$^2$/s$^2$ & $\overline{v'w'}$ & \texttt{vw}      \\
$z$ Reynolds normal stress                                                                                   & m$^2$/s$^2$ & $\overline{w'w'}$ & \texttt{ww}      \\ \hline
\end{tabular}
\end{table}

\begin{table}[]\caption{Derived feature fields available in the dataset. For the definition of $\nabla U$, $i$ is the row index, and $j$ is the column index. All fields are derived based on cell center quantities for the collocated grid arrangement in OpenFOAM, which means that $\text{trace}(\nabla U)$ may not be zero. The divergence-free velocity field imposed by the continuity equation is enforced at the cell faces, and Rhie-Chow interpolation\cite{Rhie1983} is used to handle pressure-velocity coupling on the collocated grid.}\label{tbl:fields_derived}
\centering
\small
\begin{tabular}{ccccc}
\hline
Quantity                                                                                                                                               & Units    & Symbol                     & Field name        & Expression                                                                                                                                        \\ \hline
\multicolumn{5}{c}{Features from RANS}                                                                                                                                                                                                                                                                                                                                \\   \hline    Mean velocity gradient tensor                                                                                                                          & s$^{-1}$ & $\nabla U$                 & \texttt{gradU}   & $\displaystyle \frac{\partial U_i}{\partial x_j}$                                                                                                 \\ 
Mean strain rate tensor                                                                                                                                & s$^{-1}$ & $S$                   & \texttt{S}       & $\displaystyle \tfrac{1}{2}\left(\nabla U + \nabla U ^T \right)$                                                                                  \\
Mean rotation rate tensor                                                                                                                              & s$^{-1}$ & $R$                   & \texttt{R}       & $\displaystyle \tfrac{1}{2}\left(\nabla U - \nabla U ^T \right)$                                                                                  \\
Non-dimensional strain rate tensor                                                                                                                     & -        & $\hat{S}$             & \texttt{Shat}    & $\displaystyle T_t S$                                                                                                                          \\
Non-dimensional rotation rate tensor                                                                                                                   & -        & $\hat{R}$             & \texttt{Rhat}    & $\displaystyle T_t R$                                                                                                                          \\
TKE gradient vector                                                                                                                                    & m/s$^2$  & $\nabla k$                 & \texttt{gradk}   & $\displaystyle \frac{\partial k}{\partial x_j}$                                                                                                   \\
Pressure gradient vector                                                                                                                               & m/s$^2$  & $\nabla p$                 & \texttt{gradp}   & $\displaystyle \frac{\partial p}{\partial x_j}$                                                                                                   \\
\begin{tabular}[c]{@{}c@{}}Antisymmetric tensor \\ associated with $\nabla k$\end{tabular}                                                             & m/s$^2$  & $A_k$                 & \texttt{Ak}      & $\displaystyle \begin{bmatrix}0 &-\partial_z k &\partial_y k\\ \partial_z k & 0 & -\partial_x k\\ -\partial_y k &  \partial_x k & 0\end{bmatrix}$ \\
\begin{tabular}[c]{@{}c@{}}Antisymmetric tensor \\ associated with $\nabla p$\end{tabular}                                                             & m/s$^2$  & $A_p$                 & \texttt{Ap}      & See $A_k$, replacing $k$ with $p$                                                                                                            \\
Non-dimensional $A_k$                                                                                                                                  & -        & $\hat{A}_k$           & \texttt{Akhat}   & $\displaystyle \frac{\sqrt{k}A_k}{\varepsilon}$   \\[0.01cm]                                                                                           \\
Non-dimensional $A_p$                                                                                                                                  & -        & $\hat{A}_p$           & \texttt{Aphat}   & $\displaystyle \frac{A_p}{| D\vec{U}/Dt|}$                \\[0.01cm]                                                                                  \\
Turbulent time scale                                                                                                                                   & s        & $T_t$                      & \texttt{T\_t}    & $k/\varepsilon$                                                                                                                                   \\
Kolmogorov time scale                                                                                                                                  & s        & $T_k$                      & \texttt{T\_k}    & $\displaystyle \sqrt{ \frac{\nu}{\varepsilon}}$                                                                                                 \\
Pope's 10 basis tensors                                                                                                                                & -        & $\hat{\mathcal{T}}_n$ & \texttt{Tensors} & See Pope \cite{Pope1975}                                                                                                                   \\
Pope's 5 invariants of S and R                                                                                                                         & -        & $\lambda_i$                & \texttt{Lambda}  & See Pope \cite{Pope1975}                                                                                                                   \\
\begin{tabular}[c]{@{}c@{}}47 invariants of $\{\hat{S},\hat{R},\hat{A}_k,\hat{S}_p\}$, \\ as used by Wu et al.~\cite{Wu2018}\end{tabular} & -        & $\mathcal{I}$              & \texttt{I}       & See Wu et al.~\cite{Wu2018}                                                                                                                       \\
Ratio of excess rotation to strain rate                                                                                                                & -        & -                          & \texttt{q[:,0]}  & $\displaystyle \frac{\lVert \hat{R} \rVert ^2-\lVert \hat{S} \rVert ^2}{2\lVert \hat{S}\rVert^2}$                                                                   \\ \\[0.01cm]
Wall-distance based Reynolds number                                                                                                                    & -        & -                          & \texttt{q[:,1]}  & $\displaystyle \text{min}\left( \frac{\sqrt{k}y_w}{50\nu},2\right)$                                                                                 \\ \\[0.01cm]
\begin{tabular}[c]{@{}c@{}}Ratio of turbulent time scale \\ to mean strain time scale\end{tabular}                                                     & -        & -                          & \texttt{q[:,2]}  & $\displaystyle \frac{k}{\varepsilon }\lVert S \rVert$                                                                                             \\ \\[0.01cm]

\begin{tabular}[c]{@{}c@{}}Ratio of total Reynolds stress \\ to TKE\end{tabular}                                                    & -        & -                          & \texttt{q[:,3]}  & $\displaystyle \frac{\lVert \overline{u'_i u'_j} \rVert}{k}$            \\                                                                          

Wall distance & m & $y_w$ & \texttt{wallDistance} & -\\
\begin{tabular}[c]{@{}c@{}}Material derivative of velocity field \\ (equal to convective derivative)\end{tabular} & m/s$^2$ & $\text{D}U/\text{D}t$ & \texttt{DUDt} & $\displaystyle U \cdot \nabla U$

\\ \hline
\end{tabular}
\end{table}

\begin{table}[]\caption{Derived label fields available in the dataset.}\label{tbl:labels_derived}
\centering
\begin{tabular}{ccccc}
\hline
Quantity                          & Units       & Symbol      & Field name   & Expression                                                                                                                                                                                               \\ \hline
\multicolumn{5}{c}{Labels from DNS/LES}                                                                                                                                                                                                                                                \\ \hline \\[0.01cm] 
Reynolds stress tensor            & m$^2$/s$^2$ & $\tau$ & \texttt{tau} & $\displaystyle \begin{bmatrix}\overline{u'^2} &\overline{u'v'} &\overline{u'w'}\\ \overline{u'v'} & \overline{v'^2} & \overline{v'w'}\\ \overline{u'w'}&  \overline{v'w'} &\overline{w'^2}\end{bmatrix}$ \\ \\[0.01cm] 
Turbulent kinetic energy                               & m$^2$/s$^2$ & $k$         & \texttt{k} & $\displaystyle \tfrac{1}{2}\text{trace}(\tau)$                                                                                                                                                      \\ \\[0.01cm] 
Non-dimensional anisotropy tensor & -           & $b$    & \texttt{b} & $\displaystyle \frac{\tau}{2k}-\tfrac{1}{3}I$                                                                                                                                                  \\ \\[0.01cm]  \hline
\end{tabular}
\end{table}

\end{document}